\documentclass[showpacs,prl,10pt,twocolumn,superscriptaddress,aps]{revtex4}

\usepackage{ae}
\usepackage{bm} 
\usepackage{color}
\usepackage{amsmath}
\usepackage{amssymb}
\usepackage{amsfonts}
\usepackage{graphicx}
\usepackage{slashed}
\usepackage{wasysym}

\renewcommand{\vec}[1]{\mathbf{#1}}

\newcommand{\Eqref}[1]{Eq.~\eqref{#1}}

\allowdisplaybreaks

\begin{document}

\setlength{\unitlength}{1mm}

\title{Vacuum birefringence in strong inhomogeneous electromagnetic fields}

\author{Felix Karbstein}
\affiliation{Helmholtz-Institut Jena, Fr\"obelstieg 3, 07743 Jena, Germany}
\affiliation{Theoretisch-Physikalisches Institut, Abbe Center of Photonics, \\ Friedrich-Schiller-Universit\"at Jena, Max-Wien-Platz 1, 07743 Jena, Germany}
\author{Holger Gies}
\affiliation{Helmholtz-Institut Jena, Fr\"obelstieg 3, 07743 Jena, Germany}
\affiliation{Theoretisch-Physikalisches Institut, Abbe Center of Photonics, \\ Friedrich-Schiller-Universit\"at Jena, Max-Wien-Platz 1, 07743 Jena, Germany}	
\author{Maria Reuter}
\affiliation{Helmholtz-Institut Jena, Fr\"obelstieg 3, 07743 Jena, Germany}
\affiliation{Institut f\"ur Optik und Quantenelektronik, Max-Wien-Platz 1, 07743 Jena, Germany}
\author{Matt Zepf}
\affiliation{Helmholtz-Institut Jena, Fr\"obelstieg 3, 07743 Jena, Germany}
\affiliation{Institut f\"ur Optik und Quantenelektronik, Max-Wien-Platz 1, 07743 Jena, Germany}
\affiliation{Centre for Plasma Physics, School of Mathematics and Physics, Queen's University Belfast, Belfast BT7 1NN, United Kingdom}

\date{\today}

\begin{abstract}
 Birefringence is one of the fascinating properties of the vacuum of
 quantum electrodynamics (QED) in strong electromagnetic fields.  The
 scattering of linearly polarized incident probe photons into a
 perpendicularly polarized mode provides a distinct signature of the
 optical activity of the quantum vacuum and thus offers an excellent
 opportunity for a precision test of non-linear QED. Precision tests
 require accurate predictions and thus a theoretical framework that is
 capable of taking the detailed experimental geometry into account. We
 derive analytical solutions for vacuum birefringence which include
 the spatio-temporal field structure of a strong optical pump laser
 field and an x-ray probe. We show that the angular distribution of the
 scattered photons depends strongly on the interaction geometry and
 find that scattering of the perpendicularly polarized scattered
 photons out of the cone of the incident probe x-ray beam is the key
 to making the phenomenon experimentally accessible with the current
 generation of FEL/high-field laser facilities.
\end{abstract}


\maketitle
 
\begin{figure}[h]
 \centering
  \hspace*{-3mm}\includegraphics[width=1\columnwidth]{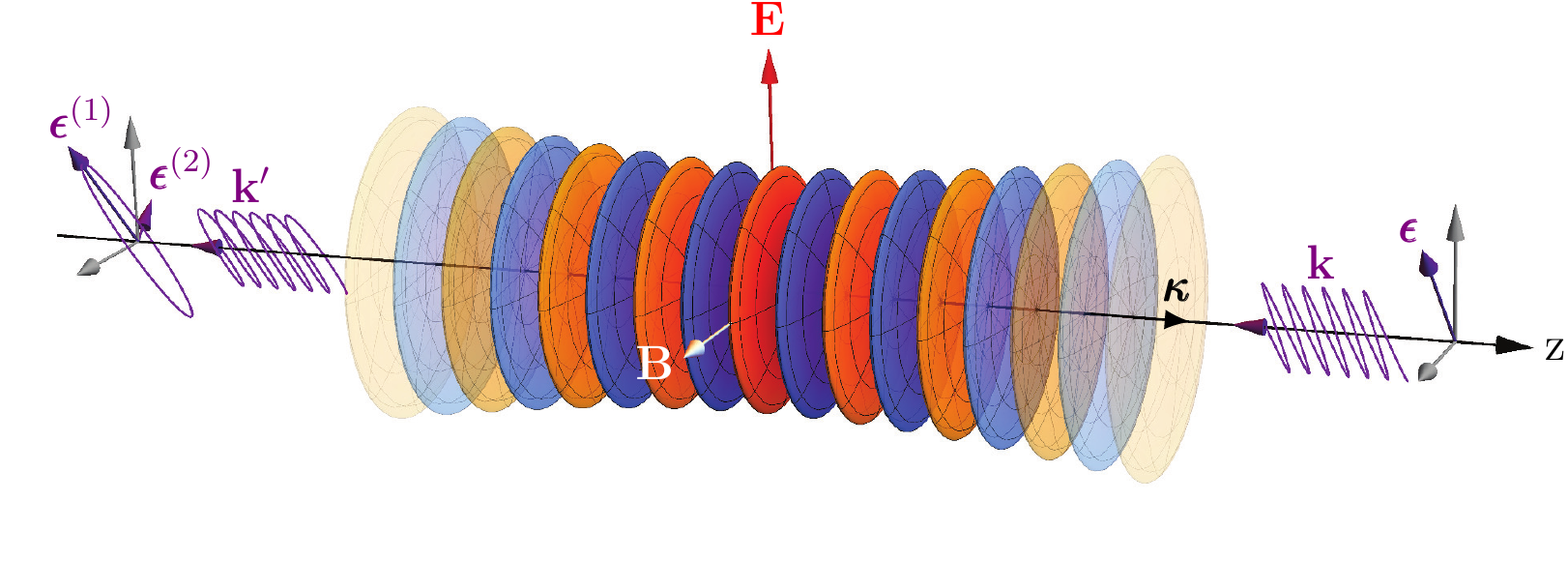}
  \vspace*{-3mm}
\caption{Sketch of the pump-probe type scenario intended to verify vacuum birefringence. A linearly polarized optical high-intensity laser pulse -- wavevector $\pmb{\kappa}$, electric (magnetic) field $\vec{E}$ ($\vec{B}$) -- propagates along the positive $\rm z$ axis. Its strong electromagnetic field couples to the charged particle-antiparticle fluctuations in the quantum vacuum, and thereby effectively modifies its properties to be probed by a counter-propagating x-ray beam (wavevector $\vec{k}$, polarization $\pmb{\epsilon}$). Vacuum birefringence manifests itself in an ellipticity of the outgoing x-ray photons (wavevector $\vec{k}'$, polarization components along $\pmb{\epsilon}^{(1)}$ and $\pmb{\epsilon}^{(2)}$).}
\label{fig:sketch}
\vspace*{-4mm}
\end{figure}

In strong electromagnetic fields the vacuum of quantum electrodynamics (QED) has peculiar properties.
Fluctuations of virtual charged particles give rise to nonlinear, effective couplings between electromagnetic fields \cite{Euler:1935zz,Heisenberg:1935qt,Weisskopf},
which, e.g., can impact and modify the propagation of light, and even trigger the spontaneous decay of the vacuum via Schwinger pair-production in electric fields \cite{Sauter:1931zz,Heisenberg:1935qt,Schwinger:1951nm}.
Even though subject to high-energy experiments \cite{Akhmadaliev:1998zz}, so far the pure electromagnetic nonlinearity of the quantum vacuum has not been verified directly on macroscopic scales.
In particular, the advent of petawatt class laser facilities has stimulated various proposals to probe quantum vacuum nonlinearities in high-intensity laser experiments; see the pertinent reviews \cite{Dittrich:2000zu,Marklund:2008gj,Dunne:2008kc,Heinzl:2008an,DiPiazza:2011tq} and references therein.
One of the most famous optical signatures of vacuum nonlinearity in strong electromagnetic fields is vacuum birefringence \cite{Toll:1952,Baier,BialynickaBirula:1970vy,Adler:1971wn,Kotkin:1996nf}, which is so far searched for in experiments using macroscopic magnetic fields \cite{Cantatore:2008zz,Berceau:2011zz}.
A proposal to verify vacuum birefringence with the aid of high-intensity lasers has been put forward by \cite{Heinzl:2006xc} (cf. also \cite{Dinu:2013gaa}), envisioning the combination of an optical high-intensity laser as pump and a linearly polarized x-ray pulse as probe; cf. also \cite{DiPiazza:2006pr} who study x-ray diffraction by a strong standing electromagnetic wave.
The basic scenario is depicted schematically in Fig.~\ref{fig:sketch}.

In the present letter, we reanalyze vacuum birefringence in a realistic all-optical experimental setup, rephrasing the phenomenon in terms of a vacuum emission process \cite{Karbstein:2014fva}, and resorting to new theoretical insights into photon propagation in slowly varying inhomogeneous electromagnetic fields \cite{Karbstein:2015cpa}.
Our study provides a new twist and perspective on the feasibility of future vacuum birefringence experiments,
emphasizing the possibility of a background free measurement by exploiting the so far completely unappreciated diffraction spreading of the outgoing signal photons in a realistic experimental setup combining a high-intensity laser system and an XFEL source as envisioned at the
Helmholtz International Beamline for Extreme Fields (HIBEF) \cite{HIBEF} at the European XFEL \cite{XFEL} at DESY.

From a theoretical perspective vacuum birefringence is most conveniently analyzed within the effective theory describing the propagation of macroscopic photon fields $A^\mu$ in the quantum vacuum. The corresponding effective action is $S_{\rm eff}[A,{\cal A}]=S_{\rm MW}[\mathbb{A}]+S_{\rm int}[A,{\cal A}]$,
where $S_{\rm MW}[\mathbb{A}]=-\frac{1}{4}\int_x \mathbb{F}_{\mu\nu}(x)\mathbb{F}^{\mu\nu}(x)$ is the Maxwell action of classical electrodynamics, with
$\mathbb{F}_{\mu\nu}$ denoting the field strength tensor of both the pump ${\cal A}^\mu$ and the probe $A^\mu$ field, i.e., $\mathbb{A}^\mu={\cal A}^\mu+A^\mu$.
The additional term $S_{\rm int}[A,{\cal A}]= -\frac{1}{2}\int_x\int_{x'}A_\mu(x)\Pi^{\mu\nu}(x,x'|{\cal A})A_\nu(x')$ encodes quantum corrections to probe photon propagation and vanishes for $\hbar\to0$.
All the effective interactions with the pump laser field
are encoded in the photon polarization tensor $\Pi^{\mu\nu}(x,x'|{\cal A})$, evaluated in the ${\cal A}^\mu$ background.
In contrast to many previous calculations, based on $\Pi^{\mu\nu}$ in the limit of constant electromagnetic fields {\it a posteriori} equipped with a qualitative pulse envelope profile,
here we account for the generically richer momentum structure in the manifestly inhomogeneous electromagnetic field of a realistic Gaussian laser pulse from the outset \cite{Karbstein:2015cpa}.
In momentum space, $\Pi^{\mu\nu}$ in inhomogeneous fields ${\cal A}_\mu(x)$ generically mediates between two independent momenta $k^\mu$ and $k'^\mu$, i.e., $\Pi^{\mu\nu}\equiv\Pi^{\mu\nu}(k',k|{\cal A})$,
This prevents a straightforward diagonalization of the probe photons' equation of motion, as possible in homogeneous fields \cite{Toll:1952,BialynickaBirula:1970vy},
where $\Pi^{\mu\nu}$ only depends on the momentum transfer $(k'-k)^\mu$ due to translational invariance.

A particular convenient way to analyze vacuum birefringence in inhomogeneous electromagnetic fields is to phrase it in terms of a vacuum emission process \cite{Karbstein:2014fva}.
Viewing the pump and probe lasers as macroscopic electromagnetic fields, and not resolving the individual photons constituting the laser beams,
the vacuum subjected to these electromagnetic fields can be interpreted as a source term for outgoing photons.
From this perspective, the induced photons correspond to the signal photons imprinted by the effective interaction of the pump and probe laser beams.
As sketched in Fig.~\ref{fig:sketch} and demonstrated below, for a linearly polarized x-ray beam brought into head-on collision with a linearly polarized optical high-intensity laser pulse, the induced x-ray signal generically encompasses photons whose polarization characteristics differ from the original probe beam.
In fact, a fraction of the induced photons is found to be polarized perpendicularly to the incident probe beam.
This can be interpreted as a signal of vacuum birefringence induced by the strong pump laser field; cf. \cite{Dinu:2013gaa}.
The outgoing x-ray beam, made up of the induced x-ray photons as well as the original probe beam which has traversed the pump laser pulse, then effectively picked up a tiny ellipticity.

We assume a linearly polarized x-ray probe beam,
$A_\nu(x)=\frac{1}{2}\frac{\cal E}{\omega}\epsilon_{\nu}(\hat{k}){\rm e}^{i\omega(\hat{k}x+t_0)-(\frac{\hat{k}x+t_0}{T/2})^2}$,
of frequency $\omega=\frac{2\pi}{\lambda_{\rm probe}}$, peak field amplitude $\cal E$ and pulse duration $T$;
$\epsilon_{\nu}(\hat{k})$ is the polarization vector of the beam and $\hat{k}^\mu=(1,\hat{\vec{k}})$, with unit wavevector $\hat{\vec{k}}$.
In momentum space, the photon current generated by the pump and probe laser fields can be expressed as
$j^\mu(k')=\frac{\sqrt{\pi}}{2}\frac{\cal E}{\omega}\frac{T}{2}\int\frac{{\rm d}\tilde\omega}{2\pi}{\rm e}^{-\frac{1}{4}(\frac{T}{2})^2(\tilde\omega-\omega)^2+it_0\tilde\omega}\,\Pi^{\mu\nu}(-k',\tilde\omega\hat{k}|{\cal A})\epsilon_{\nu}(\hat{k})$, and
the single x-ray photon emission amplitude in the macroscopic electromagnetic field $\mathbb{A}^\mu$ is given by \cite{Karbstein:2014fva}:
${\cal S}_{(p)}(k')=\frac{\epsilon_\mu^{*(p)}(\hat{k}')}{\sqrt{2\omega'}}\,j^\mu(k')$, where
the polarization vectors $\epsilon_\mu^{(p)}(\hat{k}')$, with $p\in\{1,2\}$, span the transverse polarizations of the induced photons of four-momentum $k'^\mu=\omega'(1,\hat{\vec{k}}')$.
Employing ${\cal E}=\sqrt{2\langle I\rangle}$, with the probe mean intensity given by $\langle I\rangle=J\omega$, where $J\equiv\frac{N}{\sigma T}$ is the probe photon current density, i.e., the number $N$ of incident frequency-$\omega$ photons per area $\sigma$ and time interval $T$, the differential number of induced photons with polarization $p$ determined with Fermi's golden rule,
${\rm d}^3N^{(p)}=\frac{{\rm d}^3k'}{(2\pi)^3}\bigl|{\cal S}_{(p)}(k')\bigr|^2$, can be represented as
\begin{equation}
 {\rm d}^3N^{(p)}=\frac{{\rm d}^3k'}{(2\pi)^3}\biggl|\frac{\epsilon_\mu^{*(p)}(\hat{k}')}{\sqrt{2\omega'}} {\cal M}^{\mu\nu}(-k',k|{\cal A}) \frac{\epsilon_{\nu}(\hat{k})}{\sqrt{2\omega}}\biggr|^2 J\,, \label{eq:Sp2}
\end{equation}
where we defined
\begin{multline}
 {\cal M}^{\mu\nu}(-k',k|{\cal A})=\sqrt{\pi}\,\frac{T}{2}\int\frac{{\rm d}\tilde\omega}{2\pi}\,{\rm e}^{-\frac{1}{4}(\frac{T}{2})^2(\tilde\omega-\omega)^2+it_0\tilde\omega} \\
 \times\Pi^{\mu\nu}(-k',\tilde\omega\hat{k}|{\cal A})\,. \label{eq:M}
\end{multline}

For a plane-wave probe beam, recovered in the limit $T\to \infty$, we have ${\cal M}^{\mu\nu}(-k',k|{\cal A})|_{T\to\infty}=\Pi^{\mu\nu}(-k',k|{\cal A})$.
Choosing the outgoing polarization vector $\epsilon_\mu^{(p)}(\hat{k}')$ perpendicular to the incident one,
the modulus squared term in \Eqref{eq:Sp2} can be interpreted as polarization flip probability \cite{Dinu:2013gaa}.

We assume the high-intensity laser with normalized four wave vector $\hat\kappa^\mu=(1,\vec{e}_{\rm z})$ to be linearly polarized along $\vec{e}_E=(\cos\phi,\sin\phi,0)$ (cf. Fig.~\ref{fig:sketch}, where $\phi=0$). 
The choice of the angle parameter $\phi$ fixes the directions of the electric and magnetic fields ($\vec{e}_B=\vec{e}_E|_{\phi\to\phi+\frac{\pi}{2}}$).
Moreover, $k_\perp^\mu=(0,\vec{k}-(\hat{\pmb\kappa}\cdot\vec{k})\hat{\pmb\kappa})$ and $k'^\mu_\perp=k_\perp^\mu|_{\vec{k}\to\vec{k}'}$.
For the following discussion it is convenient to turn to spherical coordinates and express the unit momentum vectors as 
$\hat{\vec{k}}=(\cos\varphi\sin\vartheta,\sin\varphi\sin\vartheta,-\cos\vartheta)$, such that $\hat{\vec{k}}|_{\vartheta=0}=-\vec{e}_{\rm z}$,
and likewise $\hat{\vec{k}}'=\hat{\vec{k}}|_{\varphi\to\varphi',\vartheta\to\vartheta'}$.
Correspondingly, the polarization vectors can be expressed as $\vec{\epsilon}_\mu(\hat{k})=(0,\vec{e}_{\varphi,\vartheta,\beta})$, with
\begin{equation}
\vec{e}_{\varphi,\vartheta,\beta}=
-\left(\begin{array}{c}
  \cos\varphi\cos\vartheta\cos\beta+\sin\varphi\sin\beta \\
  \sin\varphi\cos\vartheta\cos\beta-\cos\varphi\sin\beta \\
  \sin\vartheta\cos\beta
 \end{array}\right) , \label{eq:e_perpbeta}
\end{equation}
and $\vec{\epsilon}_\mu^{(p)}(\hat{k}')=(0,\vec{e}_{\varphi',\vartheta',\beta'})$. Without loss of generality $\vec{\epsilon}_\mu^{(1)}(\hat{k}')$ is fixed by a particular choice of $\beta'$,
and the perpendicular vector by $\vec{\epsilon}_\mu^{(2)}(\hat{k}')=\vec{\epsilon}_\mu^{(1)}(\hat{k}')|_{\beta'\to\beta'+\frac{\pi}{2}}$.

On shell, i.e., for $\hat{k}^2=\hat{k}'^2=0$, the photon polarization tensor in a linearly polarized, pulsed Gaussian laser beam is of the following structure (cf. Eq.~(16) of \cite{Karbstein:2015cpa})
\begin{multline}
 \Pi^{\mu\nu}(-k',\tilde{\omega}\hat{k}) = -\omega'\tilde\omega\frac{\alpha}{\pi}\frac{1}{45}\frac{I_0}{I_{\rm cr}}\, g(k'-\tilde{\omega}\hat{k}) \\
 \times\bigl[ 4\,(\hat k'\hat F)^\mu  (\hat k\hat F)^\nu + 7\,(\hat k'{}^*\hat F)^\mu (\hat k{}^*\hat F)^\nu \bigr]\,, \label{eq:Picrossed}
\end{multline}
where $g(k'-\tilde{\omega}\hat{k})=\int_x{\rm e}^{-i(k'-\tilde{\omega}\hat{k})x}\, g(x)$ is the Fourier transform of the normalized intensity profile of the pump laser in position space, $g(x)=I(x)/I_0$, with peak intensity $I_0$; $\alpha=\frac{e^2}{4\pi}\simeq\frac{1}{137}$, and
$I_{\rm cr}=(\frac{m^2}{e})^2\approx 4.4\cdot10^{29}\frac{\rm W}{{\rm cm}^2}$ is the critical intensity.
Here, the tensor structure is expressed in terms of $(\hat{k}\hat F)^{\mu}=\varepsilon_1^\mu(\hat{k})\cos\phi + \varepsilon_2^\mu(\hat{k})\sin\phi$ and $(\hat{k}{}^*\hat F)^{\mu}=(\hat{k}\hat F)^{\mu}|_{\phi\to\phi+\frac{\pi}{2}}$,
with $\varepsilon_1^\mu(\hat{k})=(-\hat{k}_{\rm x},\hat\kappa \hat{k},0,-\hat{k}_{\rm x})$ and $\varepsilon_2^\mu(\hat{k})=(-\hat{k}_{\rm y},0,\hat\kappa\hat{k},-\hat{k}_{\rm y})$.
The polarization dependence of the induced photon signal spanned by the two transverse polarization modes $\epsilon_\mu^{(p)}(\hat{k}')$
is encoded in the tensor structure in \Eqref{eq:Picrossed} contracted with the polarization vector of the probe beam and  $\epsilon_\mu^{*(p)}(\hat{k}')$; cf. \Eqref{eq:Sp2}.
With the above definitions we obtain
\begin{multline}
 \epsilon_\mu^{*(p)}(\hat{k}')\bigl[4(\hat{k}'\hat F)^{\mu}(\hat{k}\hat F)^{\nu} + 7(\hat{k}'{}^*\hat F)^{\mu}(\hat{k}{}^*\hat F)^{\nu}\bigr]\epsilon_{\nu}(\hat{k}) \\
 =\!(1\!+\!\cos\vartheta')(1\!+\!\cos\vartheta)[4\cos\gamma'\cos\gamma\!+\!7\sin\gamma'\sin\gamma]\,, \label{eq:Pmunu}
\end{multline}
where $\gamma=\varphi-\beta-\phi$ and $\gamma'=\varphi'-\beta'-\phi$, i.e., \Eqref{eq:Pmunu} depends on the direction ($\varphi$, $\theta$) and polarization ($\beta$) of the incident probe photons, the polarization of the pump laser beam $\phi$,
as well as the emission direction ($\vartheta',\varphi'$) and polarization ($\beta'$) of the induced photons.

Plugging Eqs.~\eqref{eq:Picrossed}-\eqref{eq:Pmunu} into Eqs.~\eqref{eq:Sp2}-\eqref{eq:M}, the differential number of induced photons can be represented as
\begin{multline}
 {\rm d}^3N^{(p)}
 =\frac{{\rm d}^3k'}{(2\pi)^3}\frac{1}{\pi}\frac{\omega'}{\omega}\Bigl(\frac{\alpha}{90}\frac{I_0}{I_{\rm cr}}\frac{T}{2}\Bigr)^2
 (1+\cos\vartheta')^2(1+\cos\vartheta)^2 \\
 \times[4\cos\gamma'\cos\gamma+7\sin\gamma'\sin\gamma]^2 \\
 \times\biggl|\int\frac{{\rm d}\tilde\omega}{2\pi}\,{\rm e}^{-\frac{1}{4}(\frac{T}{2})^2(\tilde\omega-\omega)^2+it_0\tilde\omega}\,\tilde\omega\, g(k'-\tilde\omega\hat{k})\biggr|^2 J \,.
 \label{eq:M2}
\end{multline}
The total number of induced photons is obtained upon summation over the two photon polarizations $p$, i.e., $\sum_{p}{\rm d}N^{(p)}$.
It can be inferred from \Eqref{eq:M2} by substituting
$[ 4 \cos\gamma'\cos\gamma + 7 \sin\gamma'\sin\gamma ]^2 \to [ 16  + 33 \sin^2\gamma ]$.

So far our considerations were valid for arbitrary collision geometries of the pump and probe lasers.
Subsequently, we stick to the specific scenario of counter-propagating pump and probe beams, i.e., $\hat{k}^\mu=(1,-\vec{e}_{\rm z})$,
as typically adopted in proposals for all-optical vacuum birefringence experiments \cite{Heinzl:2006xc} to maximize the overall factor of  $(1+\cos\vartheta)^2|_{\vartheta=0}=4$ in \Eqref{eq:M2}.
In this limit it is convenient to set $\varphi=0$, such that the polarization vector of the incident probe beam is $\epsilon_\mu(\hat{k})=(0,-\cos\beta,\sin\beta,0)$.
Besides, it is beneficial to ensure $t_0=0$, such that the two laser pulses have an optimal temporal overlap and the effect is maximized. 
The standard choice for scenarios aiming at the detection of vacuum birefringence in a high-intensity laser experiment is $\beta=\frac{\pi}{4}-\phi$ (cf. \cite{Heinzl:2006xc}),
implying that the polarization vector of the incident probe photons forms an angle of $\frac{\pi}{4}$ with respect to both the electric and magnetic field vectors of the pump laser; see Fig.~\ref{fig:sketch}.
This ensures an equal overlap with the two photon polarization eigenmodes featuring different phase velocities in constant crossed and plane wave backgrounds.
We will also stick to this choice here.
Note that the intensity profile of a linearly polarized, focused Gaussian laser pulse (frequency $\Omega=\frac{2\pi}{\lambda}$, beam waist $w_0$, pulse duration $\tau$ and phase $\varphi_0$), depending on the longitudinal ($\rm z$) as well as the transverse ($\rm x$, $\rm y$) coordinates, is given by \cite{Siegman}
\begin{multline}
 {\mathfrak g}(x)=\Bigl[{\rm e}^{-\frac{({\rm z}-t)^2}{(\tau/2)^2}}\, \frac{w_0}{w({\rm z})}\, {\rm e}^{-\frac{{\rm x}^2+{\rm y}^2}{w^2({\rm z})}} \\
\times\cos\Bigl(\Omega({\rm z}-t)+\tfrac{{\rm x}^2+{\rm y}^2}{w^2({\rm z})}\tfrac{\rm z}{{\rm z}_R}-\arctan\bigl(\tfrac{\rm z}{{\rm z}_R}\bigr)+\varphi_0\Bigr)\Bigr]^2 ,
\label{eq:g3}
\end{multline}
with $w({\rm z})=w_0\sqrt{1+(\frac{\rm z}{{\rm z}_R})^{2}}$ and Rayleigh range ${\rm z}_R\!=\!\frac{\pi w_0^2}{\lambda}$.

\begin{figure}[h]
 \centering
  \includegraphics[width=0.9\columnwidth]{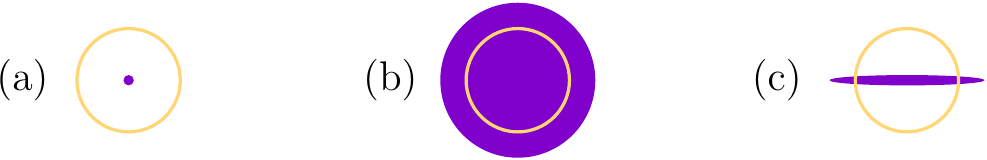}
\caption{Pictogram of the three different cases (a)-(c) considered in this letter. The circle (orange) is the cross section of the pump beam at its waist, and the filled circle/ellipse (purple) is the cross section of the probe beam.}
\label{fig:cases}
\end{figure}

In this letter, we consider three generic cases (cf. Fig.~\ref{fig:cases}), namely (a) the radius of the x-ray probe beam is significantly smaller than the beam waist of the pump laser, (b) the probe beam radius is substantially larger than the beam waist of the pump \footnote{The beam waist $w_0$ is assumed to be an adequate measure of the pump's beam radius as the intensity falls off rapidly outside the focus.}, and (c) an asymmetric x-ray beam profile which is substantially smaller than the beam waist of the pump laser in one, and larger in the other transverse direction.

To tackle case (a) theoretically, we assume that the radius of the probe beam is so tiny that basically all probe photons are propagating on the beam axis of the pump laser beam where the laser field becomes maximum, and hence do not sense the pump's transverse structure.
To this end, for $g(x)$ we adopt the on-axis profile of a focused Gaussian laser pulse, i.e., do not account for the transverse structure of the beam, such that $g(x)=\mathfrak{g}(x)|_{{\rm x}={\rm y}=0}$.
Correspondingly, the transverse momentum components remain unaffected and we have $\hat{\vec{k}}'=\hat{\vec{k}}=-\vec{e}_{\rm z}$, implying
that the photons carrying the birefringence signal will be propagating in the direction of the original probe beam, and we can set $\varphi'=0$.
This kinematic restriction requires $\omega'=|k_{\rm z}'|$, but allows for $k'_{\rm z}\neq k_{\rm z}$ and $\omega'\neq\omega$.
In the evaluation of \Eqref{eq:M2} we can then make use of the homogeneity in the transverse directions, implying $g(k'-\tilde\omega\hat{k})\sim(2\pi)^2\delta^{(2)}(k_\perp)$, and $\int_{k_{\perp}}{\rm d}^3N^{(p)}\sim J\sigma=\frac{N}{T}$.

Conversely, for case (b), where the probe photons inherently probe the transverse structure of the pump beam, such that in general ${k}'^\mu\neq{k}^\mu$, we take into account the full intensity profile of a focused Gaussian laser pulse, i.e., adopt $g(x)=\mathfrak{g}(x)$.

For case (c) we do not account for the transverse structure of the Gaussian beam along one direction, say $\rm y$, but fully take it into account in $\rm x$ direction.
Hence, we adopt $g(x)=\mathfrak{g}(x)|_{{\rm y}=0}$, implying $k'_{\rm y}=k_{\rm y}$, but generically 
${k}'^\mu\neq{k}^\mu$ for $\mu\in\{0,1,2\}$.
Here, we have $g(k'-\tilde\omega\hat{k})\sim(2\pi)\delta(k_{\rm y})$, such that $\int_{k_{\rm y}}{\rm d}^3N^{(p)}\sim JL_{\rm y}=\frac{N}{TL}$, with $L\equiv L_{\rm x}$.

As demonstrated below, in all of the cases (a)-(c) the induced x-ray photons are emitted in directions $\hat{\vec{k}}'$ very close to the propagation direction $\hat{\vec{k}}$ of the probe laser beam, and hence fulfill $\vartheta'\ll 1$.
Within at most a few tens of $\mu{\rm rad}$ the signal falls off rapidly to zero, and we have
$\vec{\epsilon}_\mu^{(p)}(\hat{k}')=\bigl(0,-\cos(\beta'-\varphi'),\sin(\beta'-\varphi'),0\bigr)+{\cal O}(\vartheta')$.
This implies that one can decompose the induced photon signal into photons polarized parallel, $\vec{\epsilon}_\mu^{(\parallel)}(\hat{k}')$ ($\beta'=\frac{\pi}{4}-\phi+\varphi'$), and perpendicular, $\vec{\epsilon}_\mu^{(\perp)}(\hat{k}')$ ($\beta'=\frac{3\pi}{4}-\phi+\varphi'$),
to the probe.
In turn, \Eqref{eq:M2} becomes $\varphi'$ independent, and $\int{\rm d}\varphi'\to2\pi$.
The term $[4\cos\gamma'\cos\gamma +7\sin\gamma'\sin\gamma]^2$ encoding the polarization dependence in \Eqref{eq:M2}, becomes $\frac{121}{4}$ ($\frac{9}{4}$) for the $\parallel$ ($\perp$) polarization mode.
The $\perp$-polarized photons constitute the birefringence signal \cite{Dinu:2013gaa}.

To obtain realistic estimates for the numbers of induced photons for a state of the art laser system, we assume the pump laser to be of the $1{\rm PW}$ class (pulse energy $W=30{\rm J}$, pulse duration $\tau=30{\rm fs}$ and wavelength $\lambda=800{\rm nm}$) focused to $w_0=1\mu{\rm m}$. 
The associated peak intensity is $I_0=2\frac{0.86\,W}{\tau\pi w_0^2}$; the effective focus area contains $86\%$ of the beam energy ($1/{\rm e}^2$ criterion).
For the x-ray probe we choose $\omega=12914{\rm eV}$ ($6457{\rm eV}$), for which the presently most sensitive x-ray polarimeter \cite{Uschmann:2014} was benchmarked. The polarization purity of x-rays of these frequencies can be measured to the level of $5.7\cdot10^{-10}$ ($2.4\cdot10^{-10}$).
In this letter we exclusively adopt $\omega=12914{\rm eV}$ as the birefringence signal is maximized for large $\omega$.
Assuming also the probe beam to be well-described as a focused Gaussian beam of waist $r$, its divergence is given by $\theta(r)=\frac{\lambda_{\rm probe}}{\pi r}$.
Neglecting diffraction and curvature effects for the probe beam in the actual calculation is nevertheless well-justified as long as ${\rm z}_{R,{\rm probe}}\gg{\rm z}_{R}$.
Measuring $r$ in units of $w_0$, i.e., $r=\rho w_0$, we have $\frac{{\rm z}_{R,{\rm probe}}}{{\rm z}_R}=\rho^2\frac{\omega}{\Omega}$,
implying that ${\rm z}_{R,{\rm probe}}\geq{\rm z}_R$ for $\rho^2\geq\frac{\Omega}{\omega}$.
Generically, we find $\frac{\Omega}{\omega}<{\cal O}(10^{-3})$.

\begin{figure}
 \centering
  \includegraphics[width=0.85\columnwidth]{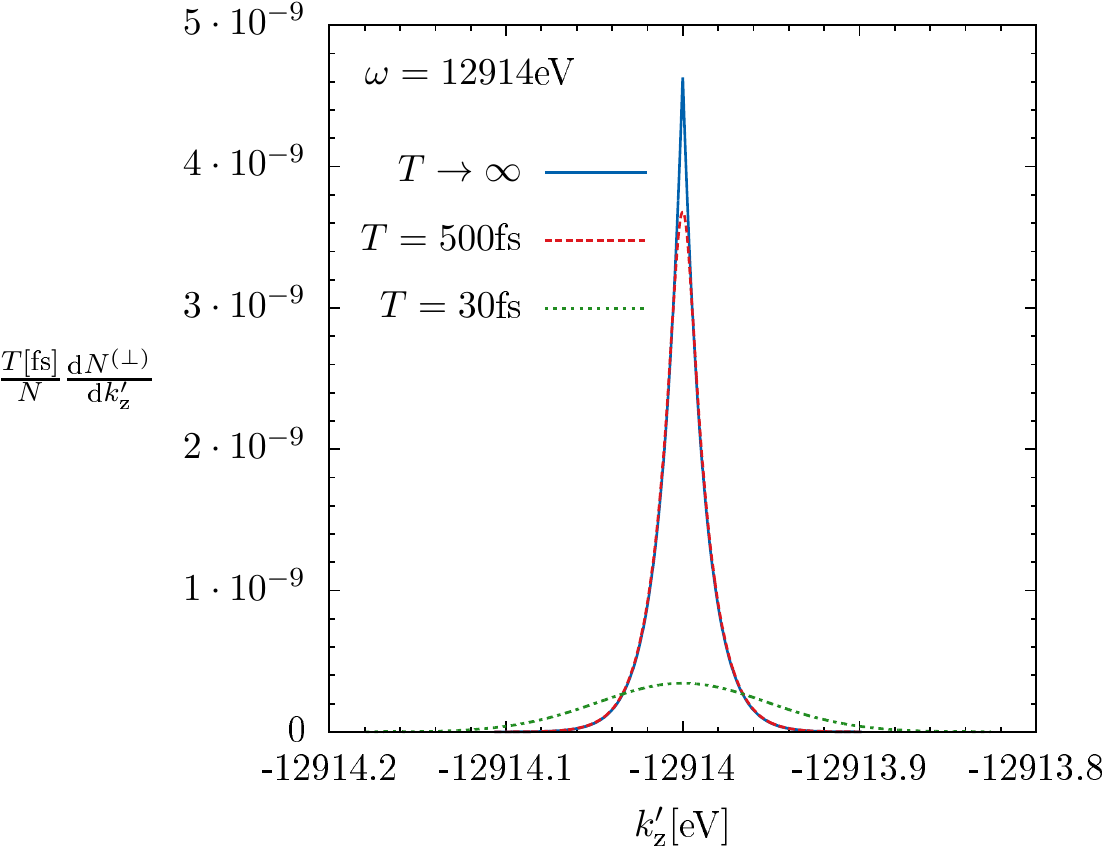}
  \hspace*{3.7cm}\begin{minipage}{5.5cm}
  \vspace*{-9.9cm}
   \includegraphics[width=0.4\textwidth]{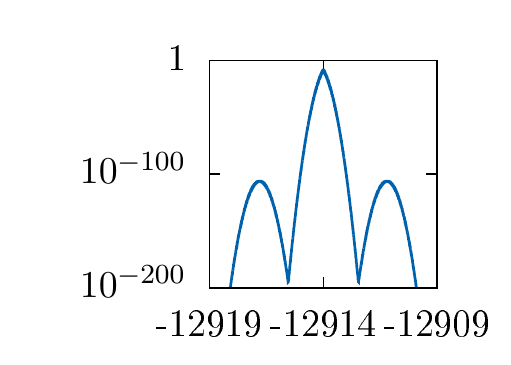}
  \end{minipage} \vspace*{-8mm}
\caption{Plot of $\frac{T}{N}\frac{{\rm d}N^{(\perp)}}{{\rm d}k'_{\rm z}}$ for case (a) as a function of $k'_{\rm z}$. The induced photon signal is peaked at $k'_{\rm z}=-\omega$ and rapidly falls off to zero. We depict results for different probe pulse durations $T$.
The inlay shows a plot of the same quantity for $T\to\infty$ ($\varphi_0=0$) over a wider frequency range, adopting a logarithmic scale. Here, the strongly suppressed contributions to be associated with frequencies $\approx \omega\pm2\Omega$ are clearly visible.}
\label{fig:dN1}
\end{figure}

We first focus on case (a), where $r\ll w_0$. In Fig.~\ref{fig:dN1} we plot $\frac{T}{N}\frac{{\rm d}N^{(\perp)}}{{\rm d}k'_{\rm z}}$. 
Upon integration over $k'_{\rm z}$ we obtain
\begin{table}[h!]
(a)\quad
 \begin{tabular}{c|c|c}
  $T$[fs] & $N^{(\perp)}/\frac{N}{T}[\frac{1}{\rm fs}]$ & $N^{(\perp)}/N$ \\
  \hline
  $\infty$ & $1.12\cdot10^{-10}$ & -- \\
  $500$ & $1.10\cdot10^{-10}$ & $2.20\cdot10^{-13}$ \\
  $30$ & $4.16\cdot10^{-11}$ &  $1.39\cdot10^{-12}$
 \end{tabular} .
\end{table}\\
Keeping all other parameters fixed, these results can be rescaled as $(\frac{W[J]}{30})^2$ to any other pump laser energy.
Note that even the maximum ratio $\frac{N^{(\perp)}}{N}\approx 1.39\cdot10^{-12}$ obtained here for $T=30{\rm fs}$ is too small to be confirmed experimentally with currently available x-ray polarization purity \cite{Uschmann:2014}
(improvement by a factor of $\gtrsim410$ required).

\begin{figure}[h]
 \centering
  \includegraphics[width=0.94\columnwidth]{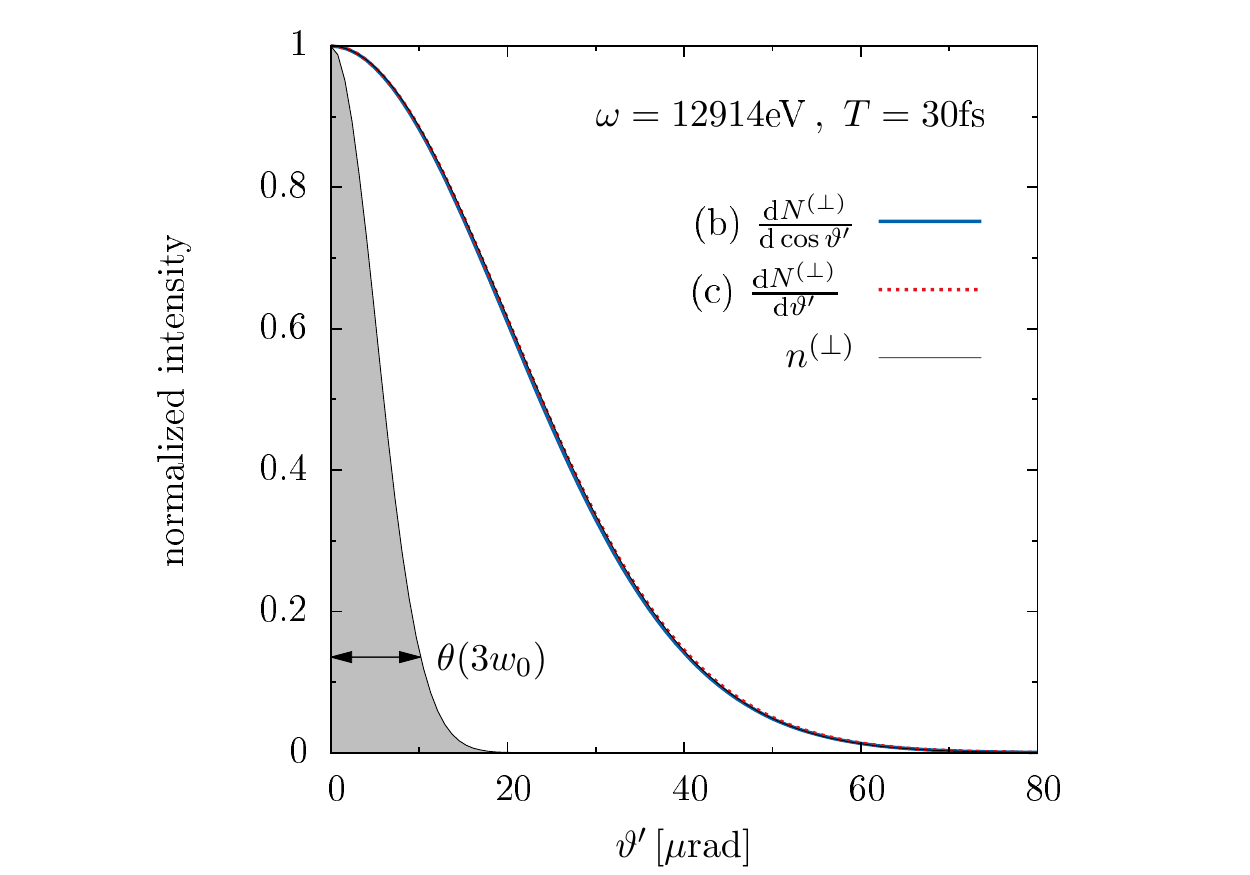}
\caption{Plot of $\frac{{\rm d}N^{(\perp)}}{{\rm d}\cos\vartheta'}\equiv2\pi\int_0^\infty{\rm}{\rm d}\omega'\omega'^2\frac{{\rm d}^3N^{(\perp)}}{{\rm d}^3k'}$ for case (b) and $\frac{{\rm d}N^{(\perp)}}{{\rm d}\vartheta'}\equiv\int{\rm d}k'_{\rm z}\int_0^\infty{\rm}{\rm d}\omega'\omega'\frac{{\rm d}^3N^{(\perp)}}{{\rm d}^3k'}$
for case (c) as a function of the emission angle $\vartheta'$.
For $T=30{\rm fs}$ the two curves basically fall on top of each other, and are well-approximated by $n^{(\perp)}(\xi)=\exp\{{-\xi(\frac{\vartheta'}{\theta(w_0)})^2}\}$ with $\xi=1.13$. This profile is compatible with naive expectations: From $N^{(\perp)}\sim I^2$ and $\omega'\approx\omega$ we might have guessed ${\rm d}N^{(\perp)}\sim n^{(\perp)}(\xi=1)$.
For comparison, the intensity profile of a Gaussian probe beam of divergence $\theta(3w_0)$ is depicted in gray.}
\label{fig:dN3}
\end{figure}

As to be expected, in scenario (b), where the probe photons sample the whole focus, the numbers of induced photons are lower.
In this case the probe beam also senses the transverse structure of the focused pump, which may result in outgoing x-ray photons with nonvanishing transverse momentum components. 
This can provide us with an additional handle to identify the induced photon signal as we can search for $\perp$-polarized photons emitted outside the divergence of the probe beam (cf. Fig.~\ref{fig:dN3}), where the demand on the polarization purity is significantly lower due to the low photon background.
Denoting the number of $\perp$-polarized photons emitted outside the beam divergence $\theta(r)$ as $N^{(\perp)}_{>\theta(r)}$, we consider probe beams of width $r=\rho w_0$ and exemplarily show results for $\rho=3$.
Identifying the cross section of the x-ray probe with $\sigma=\pi(\rho w_0)^2$, for case (b) we obtain
\begin{table}[h!]
(b)\quad
\begin{tabular}{c|c|c|c|c}
  $T$[fs] & $N^{(\perp)}/J[\frac{1}{\mu{\rm m}^2{\rm fs}}]$ & $N^{(\perp)}/N$ & $N^{(\perp)}_{>\theta(3w_0)}/N^{(\perp)}$ \\
  \hline
  $\infty$ & $1.79\cdot10^{-10}$ & -- &  $63.1\%$ \\
  $500$ & $1.52\cdot10^{-10}$ & $9.68\cdot10^{-14}/\rho^2$ & $72.1\%$ \\
  $30$ & $3.66\cdot10^{-11}$ & $3.88\cdot10^{-13}/\rho^2$ & $88.1\%$
 \end{tabular} .
\end{table}\\
Analogously, setting $L=2\rho w_0$ for case (c) we find
\begin{table}[h!]
(c)\quad
 \begin{tabular}{c|c|c|c|c}
  $T$[fs] & $N^{(\perp)}/\frac{N}{TL}[\frac{1}{\mu{\rm m}{\rm fs}}]$ & $N^{(\perp)}/N$ &  $N^{(\perp)}_{>\theta(3w_0)}/N^{(\perp)}$ \\
  \hline
  $\infty$ & $6.38\cdot10^{-11}$ & -- &  $51.1\%$ \\
  $500$ & $6.11\cdot10^{-11}$ & $1.22\cdot10^{-13}/\rho$ &  $52.6\%$ \\
  $30$ & $1.95\cdot10^{-11}$ & $6.50\cdot10^{-13}/\rho$ &  $61.7\%$ 
 \end{tabular} .
\end{table}\\
For case (b) we have $\frac{N^{(\perp)}}{N}\sim\frac{1}{\rho^2}$, while it scales as $\frac{1}{\rho}$ for case (c).
Hence, especially for large probe beam widths ($\rho>1$), case (c) is experimentally favored as it provides for the largest number of signal photons outside $\theta(\rho w_0)$.
\vspace*{0mm}\,\,\,
Finally, we demonstrate that the $\perp$-polarized photons emitted outside $\theta(r)$ can be measured with state of the art technology.
To do this, we stick to case (c) with $T=30{\rm fs}$ and $\rho=3$.
The number of $\perp$-polarized photons scattered in directions $\vartheta'\geq\vartheta_{\rm min}$ can be estimated from Fig.~\ref{fig:dN3} as $N^{(\perp)}_{>\vartheta_{\rm min}}\approx N^{(\perp)}[1-{\rm erf}(\sqrt{1.13}\frac{\vartheta_{\rm min}}{\theta(w_0)})]$, where ${\rm erf}(.)$ is the error function. Similarly, the number of probe photons outside $\vartheta_{\rm min}$ is
$N_{>\vartheta_{\rm min}}= N[1-{\rm erf}(\sqrt{2}\frac{\vartheta_{\rm min}}{\theta(3w_0)})]$.
Herefrom, we can determine the value of $\vartheta_{\rm min}$ for which the ratio $\frac{N^{(\perp)}}{N}\big|_{>\vartheta_{\rm min}}$
exceeds the polarization purity of the presently best x-ray polarimeter.
We find that this is achieved for $\vartheta_{\rm min}\geq19.50\mu{\rm rad}$.
For this choice we have $\frac{N^{(\perp)}_{>\vartheta_{\rm min}}}{N}\gtrsim 7.37\cdot10^{-14}$, such that, assuming the probe pulse to comprise $N=10^{12}$ photons and a repetition rate of $1{\rm Hz}$, we expect to detect $N^{(\perp)}_{>\vartheta_{\rm min}}\approx265$ $\perp$-polarized photons per hour.
Reducing the pulse energy of the pump to $W=3{\rm J}$, while keeping all other parameters fixed,
results in the requirement $\vartheta_{\rm min}\geq25.10\mu{\rm rad}$, for which $\frac{N^{(\perp)}_{>\vartheta_{\rm min}}}{N}\gtrsim 4.77\cdot10^{-16}$.
Due to the lower pulse energy, the repetition rate can be increased to $10{\rm Hz}$, yielding $N^{(\perp)}_{>\vartheta_{\rm min}}\approx17$ $\perp$-polarized photons per hour.

In conclusion, we have demonstrated that vacuum birefringence can be verified experimentally with state of the art technology. 
The key idea to making this phenomenon experimentally accessible is to exploit the scattering of $\perp$-polarized photons out of the cone of the incident probe x-ray beam.
For the treatment of realistic laser fields, it is computationally efficient to reformulate vacuum birefringence as vacuum emission \cite{Karbstein:2014fva}, and employ new theoretical insights into photon propagation in slowly varying inhomogeneous fields \cite{Karbstein:2015cpa}.
We are optimistic that our study can pave the way for an experimental verification of vacuum birefringence in an all-optical experiment.\\

\acknowledgments

We acknowledge support by the DFG (SFB-TR18) and EPSRC.

\onecolumngrid

\section*{Supplemental Material}

For completeness, we provide the explicit formulae used to determine the differential numbers of induced photons in this letter.
As detailed in the main text, the pump laser is assumed to be of the $1{\rm PW}$ class (pulse energy $W=30{\rm J}$, pulse duration $\tau=30{\rm fs}=45.60{\rm eV}^{-1}$, wavelength $\lambda=800{\rm nm}$ and phase $\varphi_0$)
focused to $w_0=1\mu{\rm m}=5.06{\rm eV}^{-1}$. 
The associated peak intensity is $I_0=2\frac{0.86\,W}{\tau\pi w_0^2}=8.74\cdot10^{16}{\rm eV}^4$ (the critical intensity is $I_{\rm cr}=(\frac{m^2}{e})^2=7.43\cdot10^{23}{\rm eV}^4$), the Rayleigh range ${\rm z}_R=\frac{\pi w_0^2}{\lambda}=19.91{\rm eV}^{-1}$ and
the frequency $\Omega=\frac{2\pi}{\lambda}=1.55{\rm eV}$.
The x-ray probe (frequency $\omega=12914{\rm eV}$) is assumed to deliver $N=10^{12}$ photons per pulse.
Correspondingly, the probe photon current density per area $\sigma$ and time interval $T$ is given by $J=\frac{N}{\sigma T}$.
We identify the parameter $T$ with the pulse duration of the probe laser, and $\sigma=\pi r^2$ with the cross section of the probe laser beam at its waist $r=\rho w_0$ (diameter $L=2\rho w_0$), measured in units of $w_0$.
Moreover, we set the time delay between the pump and probe laser pulses to zero, i.e., $t_0=0$, such that the two laser pulses have an optimal temporal overlap and the effect is maximized. 
For the most general situation, i.e., whenever we account for a finite probe pulse duration $T$ ($T=30{\rm fs}=45.60{\rm eV}^{-1}$ and $T=500{\rm fs}=760.00{\rm eV}^{-1}$), we limit ourselves to the dominant terms, which are independent of the pump frequency $\Omega$.
The explicit expression for the differential numbers of induced photons in the case (a) is
\begin{multline}
 ({\rm a}):\
 \frac{{\rm d}N^{(\perp)}}{{\rm d}k'_{\rm z}}
 \approx \frac{N}{T}\,\Theta(-k'_{\rm z})\frac{\omega'}{\omega}\Bigl(\frac{\alpha}{30}\frac{I_0}{I_{\rm cr}}\frac{\tau}{2}{\rm z}_R\frac{T}{2}\Bigr)^2
 \frac{1}{[(\frac{T}{2})^2+\frac{1}{2}(\frac{\tau}{2})^2]^{3}}\,{\rm e}^{-\frac{1}{2}(\frac{T}{2})^2 (|k'_{\rm z}|-\omega)^2} \\
 \times\biggl|\sum_{l=\pm1}\,
 \Bigl[\frac{1}{2}\Bigl(\frac{T}{2}\Bigr)^2\omega+\frac{1}{4}\Bigl(\frac{\tau}{2}\Bigr)^2|k'_{\rm z}|+it_0-2l{\rm z}_R\Bigr] \\
 \times{\rm e}^{\bigl(\frac{l[\frac{1}{2}(\frac{T}{2})^2(\omega-|k'_{\rm z}|)+it_0]-2{\rm z}_R}{[(\frac{T}{2})^2+\frac{1}{2}(\frac{\tau}{2})^2]^{1/2}}\bigr)^2}
 \frac{1}{2}\biggl[1+{\rm erf}\biggl(\frac{l[\frac{1}{2}(\frac{T}{2})^2(\omega-|k'_{\rm z}|)+it_0]-2{\rm z}_R}{[(\frac{T}{2})^2+\frac{1}{2}(\frac{\tau}{2})^2]^{1/2}}\biggr)\biggr]\biggr|^2\,,
\end{multline}
where  $\Theta(.)$ is the Heaviside function and ${\rm erf}(.)$ is the error function. For the other two cases considered in this letter we obtain
\begin{multline}
 ({\rm b}):\
 \frac{{\rm d}^3N^{(\perp)}}{{\rm d}\omega'{\rm d}\varphi'{\rm d}\cos\vartheta'}
 \approx J\,\frac{1}{\pi}\frac{\omega'}{\omega}\Bigl(\frac{\alpha}{480}\frac{I_0}{I_{\rm cr}}\frac{\tau}{2}{\rm z}_R w_0\frac{T}{2}\Bigr)^2
 \Bigl(\frac{1+\cos\vartheta'}{\sin\vartheta'}\Bigr)^2\,{\rm e}^{-\frac{1}{4}(w_0\omega'\sin\vartheta')^2} \\
 \times\frac{2[(\frac{2{\rm z}_R}{w_0 \omega'\sin\vartheta'})^2\omega'(1+\cos\vartheta')+\frac{1}{4}(\frac{T}{2})^2\omega+\frac{1}{8}(\frac{\tau}{2})^2\omega']^2+\frac{1}{2}t_0^2}{[2(\frac{2{\rm z}_R}{w_0 \omega'\sin\vartheta'})^2+\frac{1}{4}(\frac{T}{2})^2+\frac{1}{8}(\frac{\tau}{2})^2]^3}\\
\times{\rm e}^{-(\frac{2{\rm z}_R}{w_0 \omega'\sin\vartheta'})^2\frac{
[\frac{1}{4}(\frac{T}{2})^2+\frac{1}{8}(\frac{\tau}{2})^2]\omega'^2(1-\cos\vartheta')^2
-(\frac{T}{2})^2(\omega-\omega'\cos\vartheta')(\omega'-\omega)}{2(\frac{2{\rm z}_R}{w_0 \omega'\sin\vartheta'})^2+\frac{1}{4}(\frac{T}{2})^2+\frac{1}{8}(\frac{\tau}{2})^2}}
{\rm e}^{-\frac{(\frac{T \tau}{16})^2(\omega'-\omega)^2
+\frac{1}{2}t_0^2}{2(\frac{2{\rm z}_R}{w_0 \omega'\sin\vartheta'})^2+\frac{1}{4}(\frac{T}{2})^2+\frac{1}{8}(\frac{\tau}{2})^2}} \,,
\end{multline}
with $\vartheta'\in[0\ldots\pi)$, and
\begin{multline}
 ({\rm c}):\
\frac{{\rm d}^2N^{(\perp)}}{{\rm d}\omega'{\rm d}\vartheta'}
\approx \frac{N}{TL}\,\frac{\omega'^2}{\omega}\Bigl(\frac{\alpha}{240\pi}\frac{I_0}{I_{\rm cr}}\frac{\tau}{2}{\rm z}_R w_0\frac{T}{2}\Bigr)^2
 (1+\cos\vartheta')^2 \\
\times\biggl|\int_{0}^\infty{\rm d}s\,{\rm e}^{-\frac{1}{8}[w_0\omega'\sin\vartheta'\cosh(s)]^2}\,
 \frac{(\frac{2{\rm z}_R\,{\rm sech}(s)}{w_0\omega'\sin\vartheta'})^2\omega'(1+\cos\vartheta')+\frac{1}{8}(\frac{\tau}{2})^2\omega'+\frac{1}{4}(\frac{T}{2})^2\omega+i\frac{t_0}{2}}{[2(\frac{2{\rm z}_R\,{\rm sech}(s)}{w_0\omega'\sin\vartheta'})^2+\frac{1}{8}(\frac{\tau}{2})^2+\frac{1}{4}(\frac{T}{2})^2]^{3/2}} \\
\times{\rm e}^{-2(\frac{{\rm z}_R\,{\rm sech}(s)}{w_0\omega'\sin\vartheta'})^2\frac{[\frac{1}{8}(\frac{\tau}{2})^2+\frac{1}{4}(\frac{T}{2})^2]\omega'^2(1-\cos\vartheta')^2
-(\frac{T}{2})^2(\omega-\omega'\cos\vartheta')(\omega'-\omega)-2it_0\omega'(1+\cos\vartheta')}{2(\frac{2{\rm z}_R\,{\rm sech}(s)}{w_0\omega'\sin\vartheta'})^2+\frac{1}{8}(\frac{\tau}{2})^2+\frac{1}{4}(\frac{T}{2})^2}} \\
\times{\rm e}^{-\frac{\frac{1}{2}(\frac{T\tau}{16})^2(\omega'-\omega)^2
-it_0[\frac{1}{8}(\frac{\tau}{2})^2\omega'+\frac{1}{4}(\frac{T}{2})^2\omega]+(\frac{t_0}{2})^2}{2(\frac{2{\rm z}_R\,{\rm sech}(s)}{w_0\omega'\sin\vartheta'})^2+\frac{1}{8}(\frac{\tau}{2})^2+\frac{1}{4}(\frac{T}{2})^2}}\biggr|^2 \,,
\end{multline}
where now $\vartheta'\in[0\ldots2\pi)$.
For case (a) we exemplarily also give the result for a plane wave probe (attainable in the limit $T\to\infty$) including the subleading contributions associated with the absorption/emission of two pump laser photons of frequency $\Omega$. In this limit, the differential numbers of induced photons is
\begin{multline}
 ({\rm a})\big|_{T\to\infty}:\ 
 \frac{T}{N}\frac{{\rm d}N^{(\perp)}}{{\rm d}k'_{\rm z}}
 = \Theta(-k'_{\rm z})
 \Bigl(\frac{\alpha}{60}\frac{I_0}{I_{\rm cr}}\frac{\tau}{2}\,{\rm z}_R\Bigr)^2 |k'_{\rm z}|\,\omega\,{\rm e}^{-4{\rm z}_R|\omega-|k'_{\rm z}||}\,
 \Bigl\{{\rm e}^{-\frac{1}{4}(\frac{\tau}{2})^2(\omega-|k'_{\rm z}|)^2} \\
 +4{\rm z}_R\bigl|\omega-|k'_{\rm z}|\bigr|\Bigl[\cos(2\varphi_0)\,{\rm e}^{-\frac{1}{8}(\frac{\tau}{2})^2[(\omega-|k'_{\rm z}|)^2+(|\omega-|k'_{\rm z}||-2\Omega)^2]}
 +{\rm z}_R\bigl|\omega-|k'_{\rm z}|\bigr|\,{\rm e}^{-\frac{1}{4}(\frac{\tau}{2})^2(|\omega-|k'_{\rm z}||-2\Omega)^2}\Bigr]\Bigr\}. \label{eq:dN1}
\end{multline}


\begin{thebibliography}{10}\setlength{\itemsep}{-0.5mm}

\bibitem{Euler:1935zz} 
  H.~Euler and B.~Kockel,
  Naturwiss.\  {\bf 23}, 246 (1935).

\bibitem{Heisenberg:1935qt} 
  W.~Heisenberg and H.~Euler,
  Z.\ Phys.\  {\bf 98}, 714 (1936), 
  an English translation is available at [physics/0605038].

\bibitem{Weisskopf}
V.~Weisskopf, 
Kong.\ Dans.\ Vid.\ Selsk., Mat.-fys.\ Medd.\ {\bf XIV}, 6 (1936).

\bibitem{Sauter:1931zz} 
  F.~Sauter,
  Z.\ Phys.\  {\bf 69}, 742 (1931).

\bibitem{Schwinger:1951nm} 
  J.~S.~Schwinger,
  Phys.\ Rev.\  {\bf 82}, 664 (1951).
  
\bibitem{Akhmadaliev:1998zz} 
  S.~Z.~Akhmadaliev, {\it et al.}, 
  Phys.\ Rev.\ C {\bf 58}, 2844 (1998);
  S.~Z.~Akhmadaliev, {\it et al.}, 
  Phys.\ Rev.\ Lett.\  {\bf 89}, 061802 (2002).

\bibitem{Dittrich:2000zu} 
  W.~Dittrich and H.~Gies,
  Springer Tracts Mod.\ Phys.\  {\bf 166}, 1 (2000).

\bibitem{Marklund:2008gj} 
  M.~Marklund and J.~Lundin,
  Eur.\ Phys.\ J.\ D {\bf 55}, 319 (2009).

\bibitem{Dunne:2008kc} 
  G.~V.~Dunne,
  Eur.\ Phys.\ J.\ D {\bf 55}, 327 (2009).
  
\bibitem{Heinzl:2008an} 
  T.~Heinzl and A.~Ilderton,
  Eur.\ Phys.\ J.\ D {\bf 55}, 359 (2009).
  
\bibitem{DiPiazza:2011tq} 
  A.~Di Piazza, C.~Muller, K.~Z.~Hatsagortsyan and C.~H.~Keitel,
  Rev.\ Mod.\ Phys.\  {\bf 84}, 1177 (2012).

\bibitem{Toll:1952}
J.~S.~Toll,
Ph.D. thesis, Princeton Univ., 1952 (unpublished).

\bibitem{Baier}
R.~Baier and P.~Breitenlohner, 
{Act.~Phys.~Austriaca} {\bf 25}, 212 (1967); 
{Nuov.~Cim.~B}\ {\bf 47} 117 (1967).

\bibitem{BialynickaBirula:1970vy} 
  Z.~Bialynicka-Birula and I.~Bialynicki-Birula,
  Phys.\ Rev.\ D {\bf 2}, 2341 (1970).

\bibitem{Adler:1971wn}
  S.~L.~Adler,
  {Annals Phys.}\  {\bf 67}, 599 (1971).

\bibitem{Kotkin:1996nf} 
  G.~L.~Kotkin and V.~G.~Serbo,
  Phys.\ Lett.\ B {\bf 413}, 122 (1997).
  
\bibitem{Cantatore:2008zz} 
  G.~Cantatore [PVLAS Collaboration],
  Lect.\ Notes Phys.\  {\bf 741}, 157 (2008); 
  E.~Zavattini {\it et al.} [PVLAS Collaboration],
  Phys.\ Rev.\ D {\bf 77}, 032006 (2008);
  F.~Della Valle {\it et al.},
  New\ J.\ Phys.\ {\bf 15} 053026 (2013).

\bibitem{Berceau:2011zz} 
  P.~Berceau, R.~Battesti, M.~Fouche and C.~Rizzo,
  Can.\ J.\ Phys.\  {\bf 89}, 153 (2011);
  P.~Berceau, M.~Fouche, R.~Battesti and C.~Rizzo,
  Phys.\ Rev.\ A, {\bf 85}, 013837 (2012);
  A.~Cadene, P.~Berceau, M.~Fouche, R.~Battesti and C.~Rizzo,
  Eur.\ Phys.\ J.\ D {\bf 68}, 16 (2014).

\bibitem{Heinzl:2006xc} 
  T.~Heinzl, B.~Liesfeld, K.~U.~Amthor, H.~Schwoerer, R.~Sauerbrey and A.~Wipf,
  Opt.\ Commun.\  {\bf 267}, 318 (2006).
  
\bibitem{Dinu:2013gaa} 
  V.~Dinu, T.~Heinzl, A.~Ilderton, M.~Marklund and G.~Torgrimsson,
  Phys.\ Rev.\ D {\bf 89}, 125003 (2014);
  Phys.\ Rev.\ D {\bf 90}, 045025 (2014).
  
\bibitem{DiPiazza:2006pr} 
  A.~Di Piazza, K.~Z.~Hatsagortsyan and C.~H.~Keitel,
  Phys.\ Rev.\ Lett.\  {\bf 97}, 083603 (2006).

\bibitem{Karbstein:2014fva} 
  F.~Karbstein and R.~Shaisultanov,
  Phys.\ Rev.\ D {\bf 91}, 113002 (2015).

\bibitem{Karbstein:2015cpa} 
  F.~Karbstein and R.~Shaisultanov,
  Phys.\ Rev.\ D {\bf 91}, 085027 (2015).
  
\bibitem{HIBEF}
  cf. the HIBEF website: http://www.hzdr.de/db/Cms?pOid=35325\&pNid=3214

\bibitem{XFEL}
  cf. the XFEL website: http://www.xfel.eu

\bibitem{Siegman}
A.~E.~Siegman, \textit{Lasers}, First Edition, University Science Books, USA (1986);
B.~E.~A.~Saleh and M.~C.~Teich, \textit{Fundamentals of Photonics}, First Edition, John Wiley \& Sons, USA (1991).
  
\bibitem{Uschmann:2014}
  B.~Marx, {\it et al.},
  Opt.\ Comm.\ {\bf 284}, 915 (2011); 
  Phys.\ Rev.\ Lett. {\bf 110}, 254801 (2013).
  
\end{thebibliography}
\end{document}